\documentclass[12pt,a4paper]{article}
\usepackage{epsf}
\usepackage{graphicx}
\usepackage{natbib}
\usepackage{amsmath,amssymb}
\usepackage{url}
\usepackage{xcolor}

\usepackage{journals_joge}

\def\fracd#1#2{\frac{\displaystyle #1}{\displaystyle #2}}

\begin{document}

\title{Simple approximations of some statistical functions}
\author{Zinovy Malkin \\ Pulkovo Observatory, St. Petersburg 196140, Russia \\ e-mail: malkin@gaoran.ru}
\date{\vskip -2em}
\maketitle

\begin{abstract}
Possibilities are considered to simplify the computation of several statistical
functions used to test statistical hypotheses when processing observations:
the inverse normal distribution, the Student's $t$-distribution,
and the criterion for rejecting outliers.
For these three cases, simple approximation expressions are proposed for the quantiles
of these statistical distributions, which are accurate enough for most practical
applications.
\end{abstract}


\section*{Introduction}

Statistical computations play a primary role in processing measurements and observations.
This paper examines three functions related to testing statistical hypotheses and assessing
the statistical significance of the results obtained from processing observations.
Despite the potential of using many advanced statistical packages, proprietary mathematical
software is often used to process data.
When developing this software, it is useful to have simple algorithms with sufficient accuracy
for a given application.
This is especially important for automated and large-scale computations, which are also often
conducted under conditions of limited computing resources and high-speed data processing.

The problem of simplifying statistical computations can be solved by using approximate
algorithms based on the approximation of statistical functions.
Many formulas for approximating various statistical distributions can be found in the literature;
however, their accuracy, and therefore complexity, are typically excessive for most practical applications.
It should be noted that in these studies, the authors solve the approximation problem over the entire
range of the approximated function, which is not required in practical observation processing problems.
It is sufficient to consider a fairly narrow range of confidence probabilities 0.9--0.999,
used in the vast majority of applied problems.
This allows for significant simplification of the approximation formulas and, accordingly,
reduction in computational time while maintaining sufficient approximation accuracy for practical applications.

In the papers \citet{Malkin1993a,Malkin1993b}, the author proposed simple approximating
expressions for some statistical distributions.
Since these publications were presented very briefly and are currently
not readily available, this paper presents an expanded summary of them
with additional details and refinement of the approximating formulas.


\section{Approximating expressions}
\label{sect:approximations}

This section presents simple approximation expressions for three statistical
distributions and for several significance levels most commonly used to process
observational data.
The tables \citet{BolshevSmirnov1983} and \citet{Owen1962} were used
as the initial and test data for the approximation.

\subsection{Inverse normal distribution}
\label{sect:inverse_n}

\begin{figure}
\centering
\includegraphics[width=0.7\textwidth]{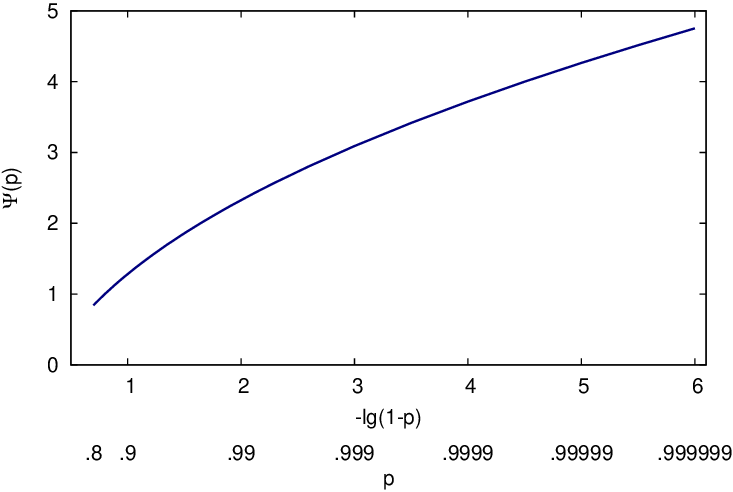}
\caption{Inverse function of the standard normal distribution.}
\label{fig:psi}
\end{figure}

The inverse function of the standard normal distribution (with zero mean and unit variance)
$x = N^{-1}(p;0.1) = \Psi(p)$ returns the value of $x$ such that $p = N(x;0.1)$.
Here $N^{-1}(x;0.1) = \Phi(x)$~ is the cumulative function of the standard normal distribution.
The form of the function $\Psi(p)$ is shown in Fig.~\ref{fig:psi}.
Strictly speaking, $x$ is the root of the equation
\begin{equation}
p = \fracd{1}{\sqrt{2\pi}} \, \int\limits_{-\infty}^{x}{e^{-\frac{t^2}{2}} dt} \,.
\label{eq:psi}
\end{equation}

The value $\Psi(p)$ is the quantile of the standard normal distribution
for the confidence probability $p$, which corresponds to a significance
level of $\alpha_1 = (1-p)$ for a one-tailed region or $\alpha_2 = 2(1-p)$
for a two-tailed region.
The significance level is often denoted by $Q$ and expressed as a percentage.
Significance levels from 0.1\% to 10\% are commonly used.
The function $\Psi(p)$ is often encountered in statistical computations,
both independently and as a component of other statistical functions,
so this section is presented in greater detail.
The values of this function can be obtained either from an exact solution
of the equation (\ref{eq:psi}) or from approximating formulas,
various versions of which are readily found in the literature.
However, such formulas are typically designed to approximate $\Psi(p)$
over the entire domain of its definition and, as a result, are quite complex.
For the main values of the confidence probability $p$ used in data
processing, a simple approximating function can be proposed that is
sufficiently accurate for practical purposes.
\begin{equation}
\Psi(p) = a_1 + a_2 \, t + a_3 \, \sqrt{t-a_4} \,, \quad t = -\ln(1-p) \,.
\label{eq:psiap}
\end{equation}

The optimal coefficients $a_1 \ldots a_4$ of formula~(\ref{eq:psiap})
depend on the given interval of $p$ values for which the best
approximation accuracy is desired.
For $p \le 0.999$, the tables of
\citet{BolshevSmirnov1983} were used, and for $p>0.999$,
the tables of \citet{Owen1962} were used.
The computation results for several $p$ intervals are presented
in Table~\ref{tab:psiap}.

\begin{table}
\centering
\caption{Results of approximation of the function $\Psi(p)$ by the formula (\ref{eq:psiap}).}
\label{tab:psiap}
\begin{tabular}{lcccccl}
\hline
Interval of $p$      && $a_1$ & $a_2$ & $a_3$ & $a_4$ & Maximum \\
                  &&       &       &       &       & error \\
\hline
$0.95 - 0.999$     && $-$0.87350465 & $-$0.02104348 & 1.61639568 & $-$0.44533427 & 0.00005 \\
$\ \, 0.9 - 0.999$      && $-$0.92337495 & $-$0.02522121 & 1.64201371 & $-$0.40330687 & 0.00010 \\
$\ \, 0.8 - 0.9999$     && $-$0.95495887 & $-$0.02695222 & 1.65576265 & $-$0.37514736 & 0.0007~ \\
$\ \, 0.8 - 0.99999$    && $-$0.92270803 & $-$0.02326696 & 1.63600922 & $-$0.39742660 & 0.0011~ \\
$\ \, 0.8 - 0.999999$   && $-$0.88998754 & $-$0.01991532 & 1.61689621 & $-$0.42100939 & 0.0012~ \\
$\ \, 0.8 - 0.99999999$ && $-$0.84935143 & $-$0.01629260 & 1.59450774 & $-$0.45174214 & 0.0024~ \\
\hline
\end{tabular}
\end{table}

Each row of the Table~\ref{tab:psiap} contains sets of coefficients $a_1 \ldots a_4$,
ensuring the best approximation accuracy over the interval specified in the first column.
The last column lists the maximum approximation error for a given interval,
which is defined as the absolute value of the difference between the values
computed by (\ref{eq:psiap}) and the exact values.
The user can select the row (set of coefficients) appropriate to their task.
The first of these intervals appears to be the most useful for practical purposes.

In practice, when implementing this algorithm in software, it is advisable
to return exact (tabular) values of $\Psi(p)$ for the most commonly used
confidence level values: 0.8, 0.9, 0.95, 0.975, 0.98,
0.99, 0.995, 0.9975, 0.999, 0.9995, and 0.9999.
It can also be noted that computing $\ln(1-p)$ does not cause a significant
loss of accuracy for the specified $p$ intervals when using double-precision arithmetic.

\citet{Brophy1985} compared several algorithms for approximating the inverse
normal distribution function, and only one of them \citep{OdehEvans1974} provides
better accuracy than the formula (\ref{eq:psiap}), namely $1 \cdot 10^{-6}$.
However, this algorithm is implemented using a much more complex formula
\begin{equation}
z = y - \fracd{c_1\,y^4+c_2\,y^3+c^3\,y^2+c_4\,y+c_5}{c_6\,y^4+c_7\,y^3+c_8\,y^2+c_9\,y+c_{10}} \,,
\quad y = \sqrt{-2\,\log(x)} \,,
\label{eq:odeh_evans}
\end{equation}
and its accuracy seems excessive for the usual practice of processing observations.
Even more accurate (and, consequently, even more complex) algorithms
for computing quantiles of a normal distribution can be found
in the literature, for example, \citet{Wichura1988},
but they are even less likely to be of practical interest.

\subsection{Student's t-distribution}
\label{sect:student}

\begin{figure}
\centering
\includegraphics[width=0.7\textwidth]{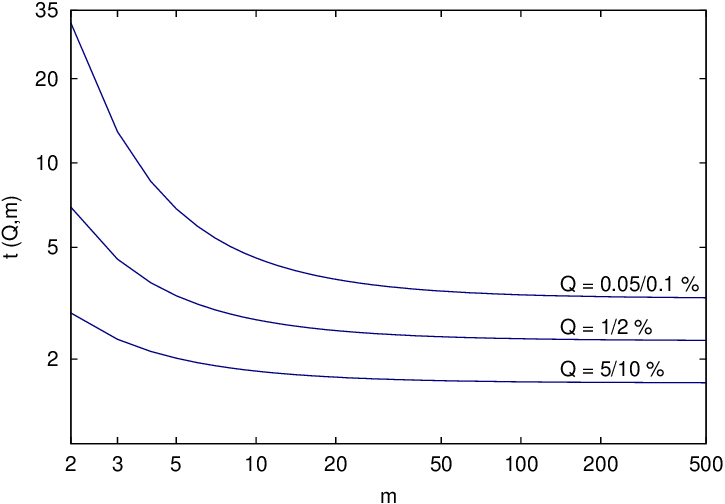}
\caption{Examples of the $t$-distribution for three significance levels:
In the graph captions, the first number corresponds to the one-tailed
region, the second to the two-tailed region.}
\label{fig:student}
\end{figure}

Another important distribution for statistical data analysis is the Student's
t-distribution, which is used, for example, to test the statistical significance
of correlation coefficients, construct confidence intervals, compare the means
of two samples, etc.
This distribution depends on one parameter--the number of degrees of freedom
$m$--and has the form shown in Fig.~\ref{fig:student}.

The number of degrees of freedom when applying tests based on the
t-distribution depends on the specific application.
For example, when testing the hypothesis that two sample means are equal,
the number of degrees of freedom is $m=n_1+n_2-2$, where $n_1$ and $n_2$
are the sample sizes.
And when testing the hypothesis that the correlation coefficient is zero,
the number of degrees of freedom is $m=n-2$.

A simple expression for computing the quantiles of the $t$-distribution,
yet providing sufficient accuracy for practical use, is:
\begin{equation}
t(Q,m) = a_1 + \fracd{a_2}{m+a_3} \,,
\label{eq:tqm}
\end{equation}
where $Q$ is the significance level, $m$ is the number of degrees of freedom.
The values of the coefficients in the formula (\ref{eq:tqm}) are given in Table~\ref{tab:student}.
The first two columns contain the significance levels as percentages for the one-tailed
and two-tailed regions.
Columns $M_1$, $M_2$, and $M_3$ contain the values of $M$ such that, for $m>M$,
the computation accuracy (the absolute value of the difference with the exact value) of
the quantiles of the $t$-distribution is no worse than 0.05, 0.01, and 0.001, respectively.

\begin{table}
\centering
\caption{Results of approximating the t-distribution using the formula (\ref{eq:tqm}).}
\label{tab:student}
\begin{tabular}{ccccccccc}
\hline
$Q_1$, \% & $Q_2$, \% & $a_1$ & $a_2$ & $a_3$ & $M_1$ & $M_2$ & $M_3$ & $t(Q,\infty)$ \\
\hline
10   & 20  & 1.2815 &  0.8483 & -0.6407 & 2 & 3 &    & 1.2816 \\
     &     & 1.2815 &  0.8476 & -0.6505 &   &   &  5 &  \\[1.2ex]
5    & 10  & 1.6448 &  1.5285 & -0.8798 & 3 & 4 &    & 1.6449 \\
     &     & 1.6448 &  1.5249 & -0.9050 &   &   &  5 &  \\[1.2ex]
2.5  & 5   & 1.9598 &  2.3848 & -1.1072 & 3 & 4 &    & 1.9600 \\
     &     & 1.9599 &  2.3759 & -1.1457 &   &   &  6 &  \\[1.2ex]
1    & 2   & 2.3259 &  3.7626 & -1.3982 & 4 & 5 &    & 2.3263 \\
     &     & 2.3263 &  3.7396 & -1.4587 &   &   &  7 &  \\[1.2ex]
0.50 & 1   & 2.5750 &  4.9793 & -1.6092 & 5 & 6 &    & 2.5758 \\
     &     & 2.5757 &  4.9356 & -1.6932 &   &   &  7 &  \\[1.2ex]
0.25 & 0.5 & 2.8055 &  6.3402 & -1.8126 & 5 & 6 &    & 2.8070 \\
     &     & 2.8068 &  6.2630 & -1.9258 &   &   &  8 &  \\[1.2ex]
0.10 & 0.2 & 3.0873 &  8.3566 & -2.0689 & 5 & 6 &    & 3.0902 \\
     &     & 3.0897 &  8.2103 & -2.2253 &   &   &  9 &  \\[1.2ex]
0.05 & 0.1 & 3.2860 & 10.0454 & -2.2535 & 6 & 6 &    & 3.2905 \\
     &     & 3.2898 &  9.8193 & -2.4492 &   &   & 10 &  \\
\hline
\end{tabular}
\end{table}

The algorithm for approximating the t-distribution described here was tested using
Table 3.2 from \citet{BolshevSmirnov1983}, which provides data for $m=1 \ldots 500$.
At the same time, it is known that for $m \rightarrow \infty$, the t-distribution tends
toward a normal distribution, so that
\begin{equation}
\begin{array}{rcl}
t(Q,\infty) &=& \Psi(1-Q/100) \quad \mbox{one-tailed region}, \\
t(Q,\infty) &=& \Psi(1-Q/200) \quad \mbox{two-tailed region}.
\label{eq:t_infinity}
\end{array}
\end{equation}
These values are given in the last column of Table~\ref{tab:student}.
On the other hand, for $m \rightarrow \infty$, the approximation formula (\ref{eq:tqm})
transforms into $t(Q,\infty) = a_1$.
Comparison of columns $a_1$ and $t(Q,\infty)$ of Table~\ref{tab:student}
shows that these values are very close to each other, especially for the variant
of the coefficients given in the second row for each $Q$.
Thus, formula (\ref{eq:tqm}) is also applicable for $m > 500$ while maintaining
the specified approximation error.

\subsection{Rejecting outliers}
\label{sect:outliers}

When processing observations, the problem of rejecting outliers
(data points that differ significantly from other data points) often arises.
One well-known procedure for statistically identifying outlier is as follows.
Let's assume that, from processing a series of observations containing
$n$ data points, the mean value $\hat x$ and its variance are obtained:
\begin{equation}
\hat x = \fracd{1}{n} \sum\limits_{i=1}^n x_i \,, \quad
s = \fracd{1}{n-1} \sum\limits_{i=1}^n v_i^2 \,, \quad
v_i = x_i-\hat x \,. \\
\label{eq:stat1}
\end{equation}

\begin{figure}
\centering
\includegraphics[width=0.7\textwidth]{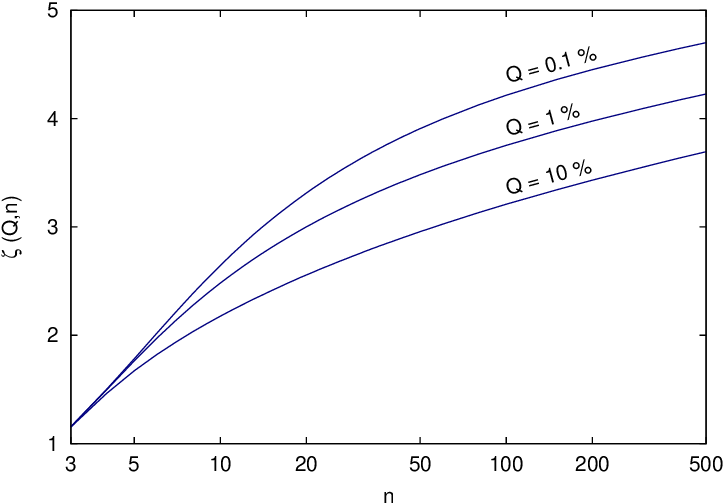}
\caption{Examples of the $\zeta(Q,n)$ distribution for three significance levels.}
\label{fig:zeta}
\end{figure}

Suppose that the $k$-th data point has the maximum residual $v_k$
and it is necessary to check whether it is an outlier to be rejected.
The rejection question is decided positively if the following
condition is met:
\begin{equation}
\frac {|v_k|}{s} > \zeta (Q,n) \,,
\label{eq:condition}
\end{equation}
where $Q$ is the significance level.
The form of the function $\zeta (Q,n)$ for three significance levels
is shown in Fig.~\ref{fig:zeta}.

This Section presents expressions for the approximate computation
of the statistic $\zeta (Q,n)$ that are simple enough and accurate
for practical applications.
For the approximation, Table 4.8c from \citet{BolshevSmirnov1983}
was used, with two differences.
First, the table was extended to $n=500$ using the formula
\citet{BolshevSmirnov1983}, p. 60, and the approximation
$\Psi(p)$ described in section \ref{sect:inverse_n}.
Second, Table 4.8c from \citet{BolshevSmirnov1983} is given for the case
of computing the variance as $s=\fracd{1}{n} \sum v_i^2$ {\Large \strut}.
Therefore, it was updated for the case of computing the sample variance
as $s=\fracd{1}{n-1} \sum v_i^2$ {\LARGE \strut} by multiplying the values
from \citet{BolshevSmirnov1983} by $\sqrt\fracd{n-1}{n}$ {\Large \strut}.

The approximation of the function $\zeta (Q,n)$ is performed according
to the formula
\begin{equation}
\zeta (Q,n) = \Psi(p) \cdot \left( a_1 + a_2 \, n + \fracd{a_3}{a_4+n} \right) \,,
\label{eq:zeta}
\end{equation}
where $p = 1-Q/200n$, and $Q$ is expressed as a percentage.
The values of the coefficients $a_1 \ldots a_4$ for different values
of $Q$ are given in Table~\ref{tab:zeta}.
For each $Q$, the table provides two variants of the coefficients.
The top row shows the optimal coefficients for $n=6 \ldots 500$,
and the bottom row shows them for $n=6 \ldots 100$.
The approximation error (the absolute value of the difference between the
approximate values computed using (\ref{eq:zeta}) and the exact values)
is less than $0.007$ in the first case and $0.003$ in the second.

\begin{table}
\centering
\caption{Results of approximation of the function $\zeta (Q,n)$ by the formula (\ref{eq:zeta}).}
\label{tab:zeta}
\begin{tabular}{cccccc}
\hline
$Q$, \% & n & $a_1$                  & $a_2$                   & $a_3$      & $a_4$ \\
\hline
$10$   & $6 \ldots 500$ & $9.88545 \cdot 10^{-1}$ & $2.01128 \cdot 10^{-5}$ & $-1.68729$ & $1.63739$ \\
       & $6 \ldots 100$ & $9.81392 \cdot 10^{-1}$ & $9.79867 \cdot 10^{-5}$ & $-1.51368$ & $0.96360$ \\[1ex]
$5$    & $6 \ldots 500$ & $9.88424 \cdot 10^{-1}$ & $2.04021 \cdot 10^{-5}$ & $-1.99297$ & $1.45970$ \\
       & $6 \ldots 100$ & $9.81622 \cdot 10^{-1}$ & $9.41882 \cdot 10^{-5}$ & $-1.82875$ & $0.93060$ \\[1ex]
$2$    & $6 \ldots 500$ & $9.88432 \cdot 10^{-1}$ & $2.03774 \cdot 10^{-5}$ & $-2.41798$ & $1.52820$ \\
       & $6 \ldots 100$ & $9.81751 \cdot 10^{-1}$ & $9.28241 \cdot 10^{-5}$ & $-2.25551$ & $1.09606$ \\[1ex]
$1$    & $6 \ldots 500$ & $9.88913 \cdot 10^{-1}$ & $1.90686 \cdot 10^{-5}$ & $-2.76375$ & $1.72925$ \\
       & $6 \ldots 100$ & $9.82771 \cdot 10^{-1}$ & $8.45343 \cdot 10^{-5}$ & $-2.61076$ & $1.36652$ \\[1ex]
$0.5$  & $6 \ldots 500$ & $9.89111 \cdot 10^{-1}$ & $1.88814 \cdot 10^{-5}$ & $-3.10307$ & $1.95188$ \\
       & $6 \ldots 100$ & $9.83396 \cdot 10^{-1}$ & $7.86601 \cdot 10^{-5}$ & $-2.95706$ & $1.63688$ \\[1ex]
$0.2$  & $6 \ldots 500$ & $9.90087 \cdot 10^{-1}$ & $1.63353 \cdot 10^{-5}$ & $-3.58659$ & $2.37111$ \\
       & $6 \ldots 100$ & $9.85744 \cdot 10^{-1}$ & $5.91809 \cdot 10^{-5}$ & $-3.46890$ & $2.14099$ \\[1ex]
$0.1$  & $6 \ldots 500$ & $9.90843 \cdot 10^{-1}$ & $1.44077 \cdot 10^{-5}$ & $-3.95667$ & $2.71055$ \\
       & $6 \ldots 100$ & $9.87049 \cdot 10^{-1}$ & $5.12540 \cdot 10^{-5}$ & $-3.85096$ & $2.51786$ \\
\hline
\end{tabular}
\end{table}


\section{Conclusion}
\label{sect:conclusion}

This paper presents several simple expressions for approximating three statistical
distributions used in processing observational data,
in particular, for testing statistical hypotheses.
The proposed algorithms may be inferior in accuracy to other, more sophisticated
approximation methods, but their advantage is their computational simplicity,
and therefore their speed, which can be useful, and sometimes critical,
for large-scale computations, especially in real time.

Of course, the author does not claim to have an optimal solution to the stated problem
of simplifying and accelerating statistical computations and hopes that interested readers
and users will be able to propose more successful approximation solutions
for both those considered in this paper and other statistical functions.

Several Fortran routines implementing the algorithms discussed in this paper are placed
in the Pulkovo Observatory website (\url{https://www.gaoran.ru/english/as/soft/}).


\clearpage
\bibliography{malkin_statistics.bib}

\begin{thebibliography}{7}
\providecommand{\natexlab}[1]{#1}
\providecommand{\doi}[1]{doi:\discretionary{}{}{}#1}
\providecommand{\url}[1]{{#1}}
\providecommand{\eprint}[2][]{\url{#2}}

\bibitem[{{Bol'shev} and {Smirnov}(1983)}]{BolshevSmirnov1983}
{Bol'shev} LN, {Smirnov} NV (1983) {Tables of mathematical statistics}. Moscow:
  Nauka (in Russian)

\bibitem[{{Brophy}(1985)}]{Brophy1985}
{Brophy} AL (1985) {Approximation of the inverse normal distribution function}.
  Behavior Research Methods, Instruments, \& Computers 17:415--417

\bibitem[{{Malkin}(1993{\natexlab{a}})}]{Malkin1993a}
{Malkin} ZM (1993{\natexlab{a}}) {On rejecting outliers}. Sov Astron Circ
  1555:33--34 (in Russian)

\bibitem[{{Malkin}(1993{\natexlab{b}})}]{Malkin1993b}
{Malkin} ZM (1993{\natexlab{b}}) {Simple computation of t-distribution
  percentage points}. Sov Astron Circ 1554:43 (in Russian)

\bibitem[{{Odeh} and {Evans}(1974)}]{OdehEvans1974}
{Odeh} RE, {Evans} JO (1974) {Algorithm AS 70: The percentage points of the
  normal distribution}. Journal of the Royal Statistical Society Series C
  (Applied Statistics) 23:96--97

\bibitem[{{Owen}(1962)}]{Owen1962}
{Owen} DB (1962) {Handbook of statistical tables.} Pergamon Press

\bibitem[{{Wichura}(1988)}]{Wichura1988}
{Wichura} MJ (1988) {Algorithm AS 241: The percentage points of the normal
  distribution}. Journal of the Royal Statistical Society Series C (Applied
  Statistics) 37:477--484

\end{thebibliography}
\bibliographystyle{joge}

\end{document}